\documentstyle[emulateapj]{article}
\newcommand{\etal}{{ et~al.}\ }

\begin{document}
%\twocolumn[

\title{The Color Distributions of Globular Clusters in Virgo Elliptical Galaxies}
\author{Eric H. Neilsen, Jr., and Zlatan I. Tsvetanov}
\affil{Johns Hopkins University, Department of Physics and Astronomy, Baltimore, MD 21218}
\authoremail{neilsen@pha.jhu.edu}

\begin{abstract}
This {\it Letter} presents the color distributions of the globular
cluster (GC) systems of 12 Virgo elliptical galaxies, measured using
data from the {\it Hubble Space Telescope}. Bright galaxies with large
numbers of detected GC's show two distinct cluster populations with
mean $V-I$ colors near 1.01 and 1.26. The GC population of M86 is a
clear exception; its color distribution shows a single sharp peak near
$V-I=1.03.$ The absence of the red population in this galaxy, and the
consistency of the peak colors in the others, may be indications of
the origins of the two populations found in most bright elliptical
galaxies.
\end{abstract}

\keywords{galaxies: star clusters -- globular clusters: general}
%]

\section{Introduction}

The first detection of bimodality in the globular cluster system of an
elliptical galaxy was made by \markcite{zepf93} Zepf \& Ashman (1993)
using the photometry of the M49 globular cluster (GC) system taken by
\markcite{courure91} Couture \etal (1991). Better data with larger
samples of clusters subsequently confirmed the detection
(\markcite{zepf95} Zepf \etal 1995) and added several galaxies to the
list of galaxies with detected bimodal distributions. Additional
studies (\markcite{geisler96} Geisler \etal 1996, \markcite{neilsen98}
Neilsen, Tsvetanov \& Ford 1998) show that the radial color gradients
in the mean color of the globular clusters of M49 and M87 can be
attributed to variations in the spatial distributions of the clusters
contributing to each peak.

This spatially varying bimodal distribution was predicted by the
merger model of \markcite{ashman:92} Ashman \& Zepf (1992), in which
the blue clusters are those of spiral galaxies which merged to form
the elliptical, and the red clusters formed during the
mergers. \markcite{cote98} Cote \etal (1998) have proposed an
alternate model in which only the red population is truly associated
with the host galaxy, while the blue clusters were formed in other
(smaller) members of the galaxy cluster and were captured during
mergers or stripped from their original hosts through
interactions. Under this model, the red clusters are metal rich
because the large mass of the galaxy prevents metals from
escaping. The blue clusters now follow the gravitational potential of
the cluster as a whole, which is strongly centered on the central
giant ellipticals.

There are serious objections to both models. The Cote \etal (1998) model has
difficulty explaining the rotation of the GC system of M87, seen by
\markcite{kissler-patig98} Kissler-Patig \& Gebhardt (1998) in the 
data set of \markcite{cohen98} Cohen, Blakeslee, \& Ryzhov
(1998). Furthermore, the stellar metallicity of the halos of
elliptical galaxies is higher than that expected by stripping models
\markcite{harris98} (Harris, Harris \& McLaughlin 1998).
The Ashman \& Zepf (1992) model implies a correlation between the
relative numbers of red and blue clusters and specific frequency in
globular cluster systems, which is not seen.

By obtaining high precision color measurements of large, well defined
samples of GC's in elliptical galaxies located in different parts of a
galaxy cluster, further tests of these models become possible. For
both the merger model and the tidal stripping model, one would expect
the color of the blue population to be roughly the same in all
galaxies. In the tidal stripping model, the size of the galaxy is
expected to determine the location of the red peak, so a correlation
between the color of the red peak and the luminosity of the galaxy is
to be expected. (This correlation may be masked if a significant
fraction of the stars were obtained in mergers.) However, in the
merger model the shape and location of the color peak is dependent on
the details of the merger history, and one would expect significant
variation among the sample.

\section{Observations, Reduction and Analysis}

The {\it Hubble Space Telescope} archive provided all of the data used
in this study. The sample includes those elliptical galaxies in the
Virgo cluster for which there are WFPC-2 observations with images in
filters which approximate the $V$ and $I$ bands (F814W and F555W or F606W,
respectively), and which are deep enough to see a significant fraction
of the globular clusters in a galaxy at 15-20 Mpc.
Table~\ref{table:data} lists the basic parameters of the data taken.

The raw data were calibrated using the WFPC2 pipeline procedure. The
different exposures were then aligned, cosmic ray rejected, and
combined using standard software.

The custom program {\sc SBFtool} was used both to determine the
amplitude of the surface brightness fluctuations and to detect,
classify, and perform photometry on candidate objects in the
images. First, we model the galaxy and sky background using a
combination of ellipse fitting or spatial frequency filtering
techniques. We mark groups of four or more adjacent pixels
significantly brighter than the model and noise as potential object
pixels. Sets of pixels consistent with the shape of a PSF or distant
globular cluster convolved by a PSF form the set of candidate globular
clusters.

Because globular clusters at the distance of Virgo are partially
resolved in {\it HST} WFPC-2 images, simple aperture photometry
becomes complicated; the aperture correction will vary from object to
object. Instead, we fit the data to a library of model GC's
constructed using King profiles and PSF's created using {\sc tinytim}
(Krist~1997), covering a range of both core and tidal radii. (For
similar approaches to photometry on partially resolved globular
clusters, see \markcite{grillmair99} Grillmair \etal (1999) and
\markcite{holtzman96} Holtzman \etal (1996).) In central, high signal
to noise $(S/N > 30)$ pixels, where errors in the model fitting become
significant compared to the noise, the flux is measured directly. The
flux outside these pixels is estimated using the best fit model.

The remaining catalog still contains contaminating objects,
particularly background galaxies. The removal of objects with colors
outside of the range $0.5 < V-I < 1.5$ (where practically all globular
clusters fall), highly asymmetric objects, and objects where the best
fitting model provided a poor reduced $\chi^{2}$ fit significantly
reduces the contamination. For the study of the color distribution, we
consider only objects with signals greater than 3000 e$^-$ in each
filter. These objects are significantly above the detection threshold
in both filters, and the uncertainty in the color of fainter objects
due to counting statistics alone is significant. The bright magnitude
cutoff reduces the smoothing of the color distribution due to
measurement error, ensures that the catalog is complete down to a well
defined limit, and minimizes contamination due to remaining background
galaxies. In all data sets except NGC~4365, NGC~4660, and NGC~4458,
this limit is fainter than the expected peak of the GC magnitude
distribution. The limit in the M86 data set is close to the expected
peak of the GC distribution.

The procedure outlined by \markcite{holtzman95} Holtzman \etal (1995)
guided the conversion of the measured flux in the {\it HST} filters to
standard $V$ and $I$ magnitudes.

\section{Results}

\begin{deluxetable}{lcccccccccc}
\tablecaption{Globular cluster, host galaxy, and data properties \label{table:data}}
\tablehead{
 & & \colhead{Host Galaxy} & & \multicolumn{5}{c}{GC Population} &  \multicolumn{2}{c}{exp. time (s)} \\
\colhead{Galaxy} & \colhead{M} & \colhead{Type} & $B-V$ & $m_{V,lim}$ & N$_{GC}$ &\colhead{p} & \colhead{Mode 1} & \colhead{Mode 2} & \colhead{F555W}  & \colhead{F814W}  }
\startdata
NGC~4472 (M49)      & -21.6 & E2 & 0.96 & 24.1 &  239 & 0.000 & 0.99 & 1.25 &  1800 &  1800 \nl
NGC~4406 (M86)      & -21.5 & E3 & 0.93 & 24.4 &  118 & 0.563 & 1.02 & 1.15 &  1500 &  1500 \nl
NGC~4365 & -21.5 & E3 & 0.96 & 24.3 &  170 & 0.190 & 1.04 & 1.23 &  2200 &  2300 \nl
NGC~4649 (M60)      & -21.4 & E2 & 0.97 & 24.3 &  276 & 0.000 & 1.02 & 1.27 &  2100 &  2500  \nl
NGC~4486 (M87)      & -21.4 & E0 & 0.96 & 24.4 &  616 & 0.000 & 0.98 & 1.22 &  2400 &  2400  \nl
NGC~4552 (M89)      & -20.3 & E0 & 0.98 & 24.4 &  152 & 0.000 & 1.05 & 1.29 &  2400 &  1500  \nl
NGC~4621 (M59)      & -20.2 & E5 & 0.94 & 23.5 &  102 & 0.286 & 1.06 & 1.22 &  1050 &  1050 \nl
NGC~4473 & -20.0 & E5 & 0.96 & 24.1 &  107 & 0.075 & 0.99 & 1.20 &  1800 &  2000 \nl
NGC~4660 & -19.4 & E6 & 0.92 & 23.5 &  41  & 0.162 & 0.99 & 1.15 &  1000 &  850  \nl
NGC~4478 & -18.9 & E2 & 0.91 & 25.4 &  64  & 0.037 & 0.99 & 1.29 & 16800\tablenotemark{a} &  16500 \nl
NGC~4458 & -18.8 & E0-1 & 0.86 & 23.7 & 17 & 0.000 & 0.83 & 1.16 &  1200 &  1040 \nl
NGC~4550 & -18.4 & SB0 & 0.88 & 23.7 & 25  & 0.099 & 0.97 & 1.30 &  1200 &  1200 \nl 
\enddata
\tablenotetext{a}{F606W}
\end{deluxetable}

\begin{figure*}
\plotone{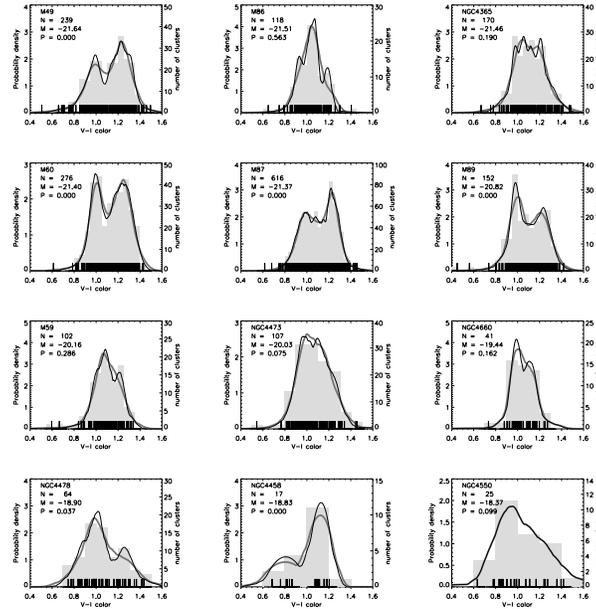}
\caption{Each plot shows the color distribution of globular clusters,
together with the number of clusters used to create the distribution,
the absolute magnitude of the parent galaxy based on the apparent
magnitude as reported in RC3 and surface brightness fluctuations
distance measurements for that galaxy, and the $p$ statistic from the
KMM test. The filled grey area shows the histogram, and the lines show
the distribution as estimated by a variable width Epanechnikov kernel
with reference kernel widths which under-smooth (thin) and over-smooth
(thick) the distribution. The ticks along the bottom represent the
measured colors of the clusters in the sample.
\label{figure:colordists}}
\end{figure*}

Figure~\ref{figure:colordists} displays the color distributions of
each galaxy. The light grey shaded area shows the distribution using a
traditional histogram. The histogram is not the ideal representation
of the data; bin sizes narrow enough to detect fine structure will
also display significant noise, and the choice of phase can also have
a significant effect on the appearance.  Simonoff (1996) describes and
compares a variety of alternatives for estimating the underlying
probability distribution, including the variable width Epanechnikov
kernel.

For any given value, one of the simplest ways to measure the density
of points at that value is to count the number of points within
some distance $h$ of that value:
\[ \hat{f}(x) =  \sum_{i=1}^{n} \frac{1}{h} K(\frac{x-x_{i}}{h})
\] where \[
 K(u) = \left\{ \begin{array}{ll} \frac{1}{2} & \mbox{if $-1 < u \leq
		1$,}\\ 0 & \mbox{otherwise.}  \end{array} \right.\]

The appropriate choice of $h$ is a function both of the form of the
underlying distribution and the density of data points; smaller values
of $h$ are warranted when there are a larger number of data
points. Use of an unnecessarily large value for $h$ will result in an
overly smooth estimate of $f(x)$. One can accommodate the varying
density of data points by substituting $h(x) = h_{\nu} \times
f(x_i)^{-\frac{1}{2}}$ for $h$. Clearly, an estimate of $f(x)$ must be
made to apply this method, but an iterative process beginning with a
crude (e.g., uniform) estimate provides stable results in few
iterations.  A second improvement that can be made is in the choice of
the function $K(u)$. It can be shown that an estimate made using the
Epanechnikov kernel,
\[ K(u) = \left\{ \begin{array}{ll} \frac{3}{4}(1-u^2) & \mbox{if $-1 < u \leq 1$,}\\ 0 & \mbox{otherwise,} \end{array} \right.\] 
minimizes the mean integrated square difference between $f(x)$ and
$\hat{f}(x)$, provided that $f^{\prime \prime}$ is continuous,
$f^{\prime \prime \prime}$ is square integrable, and $K(u) \geq
0$ (Simonoff 1996).

In figure~\ref{figure:colordists}, we present the color distributions
using a simple histogram and approximations made using two values of
reference kernel width, $h_\nu$. The thick line is the smoothing using
a reference kernel width that would be optimal for a Gaussian
distribution with a standard deviation equal to that of the data, and
estimated using a constant kernel width. This value will over-smooth
the data when applied with a variable kernel width and $\hat{f}(x)<1$,
particularly if the distribution is not Gaussian; features seen in
this smoothing are very likely to be real. A Kolmogorov-Smirnov (K-S)
test comparing this smoothed distribution to the data confirms that it
is significantly over-smoothed. The thin line shows the smoothing
using a kernel width such that the smoothed curve can be excluded by a
K-S test at the 50\% level, giving an indication of what the true
distribution may be. However, features seen in this line cannot be
reguarded as having been reliably detected.

The KMM algorithm (Ashman, Bird, \& Zepf 1994) provides a statistical
test for comparing the likelihood of the underlying distribution being
a single or double Gaussian. The KMM algorithm returns a likelihood
ratio test statistic, which is a measure of the improvement in the fit
of a two Gaussian model over a single Gaussian one. From this we
calculate the $p$ value, the probability of measuring this statistic
from a single Gaussian distribution. Low $p$ values reject the
hypothesis that the examined distribution resulted from single
Gaussian distribution. They do not necessarily reject other (possibly
unimodal) models for the distribution, however. Because the presence
of contaminating objects outside the main distribution significantly
reduces the effectiveness of the KMM algorithm, we have removed
objects with colors far from the main distribution (which are probably
contaminating background galaxies) from our sample before applying the
KMM algorithm to our data (see Ashman \etal 1994 for a more complete
discussion of the effects of such a truncation).

Table~\ref{table:data} presents the various physical properties of
each galaxy, including the absolute $B$ magnitude (calculated using SBF
distances and RC3 apparent magnitudes), $B-V$ color, and Hubble type; the $V$
magnitude cutoff and the total number of clusters considered in the
color distribution; and the $p$ value statistic and distribution
locations from the KMM algorithm (mode~1 and mode~2).

\section{Discussion}

For four of the eight data sets where a reasonably large number of
clusters have been detected $(N > 100)$, two peaks are clearly
identifiable even where the data are over-smoothed. Furthermore, in
each of these four cases the locations of the peaks are consistently
near $V-I=1.01$ and $V-I=1.26$. NGC~4365, NGC~4473, and M59 show
single peaks, broader than the individual peaks in galaxies with
bimodal distributions; the data are consistent with the base
distribution being either a single broad peak or two peaks to close to
be resolved. Although a single Gaussian distribution for these
galaxies cannot be excluded, the relatively low $p$ value and best
peak values (for the two Gaussian fit) close to 1.01 and 1.26 are
suggestive of bimodality. M86 features a large population of globular
clusters, but the distribution appears smoothly unimodal with a peak
of $V-I \simeq 1.03$. The width and color of the peak are comparable to
the width and color of the blue peak in the bimodal galaxies. The
difference seems to be its lack of a red peak. In other properties,
such as its luminosity and X-ray emission, it is similar to other
bright galaxies in the sample. The only other remarkable feature is
that M86 is bluer than any of the bright galaxies. This
may indicate that the processes which result in a detectable red peak
in the GC population also redden the star light of the galaxy as a
whole.  We do not emphasize the remaining four populations $(N < 100)$
both because of the reduced overall statistics and the fact that the
small number of clusters increases the effect of contamination by
background galaxies on the overall appearance of the distribution.

While the positions of the modes of the color distribution appear
uniform, the relative contribution of each peak to the distribution
does not. The lack of a red peak in M86, and the strength of the red
peak in the remainder of the bright galaxies, is a dramatic
illustration of this variation.

In the model where the red population if supposed to from with the
host galaxy, and the blue population is collected from neighbors, the
mass of the galaxy is reguarded as the cause for the high metallicity
of the clusters which originated with the host galaxy, so one expect a
significant variation in the position of the red peak with galaxy
mass. Our data set shows no such trend, although the range of
magnitudes may be too small for such a trend to be detected. Why the
red peak should show such consistency is equally unclear in the merger
model of formation, as the distribution of red clusters should vary
depending in the details of the merger history of the galaxy. It is
possible, though, that the breadth of the red peak hides multiple
populations formed though mergers, which typically result in a similar
overall peak.

\acknowledgments
We would like to thank Holland Ford and Patrick Cote for useful advice and discussions,
and the referee for comments which improved this presentation. Support
for this work was provided by NASA through grant GO-7543 from the
Space Telescope Science Institute, which is operated by AURA, Inc.,
under NASA contract NAS5-26555.

\end{document}